%Paper: hep-th/9209050
%From: <KPHY76@klio.rhrk.uni-kl.de>
%Date: Mon, 14 Sep 92 22:38:40 CET

\magnification1200
\hsize15true cm

\centerline{   EXTRINSIC GEOMETRY SENSITIVE BOSONIC STRING THEORY}
\bigskip
\centerline{  Al. R. Kavalov}
\bigskip
\centerline{                 Yerevan Physics Institute,   }
\bigskip
\centerline{Alikhanyan Brothers St.2, Yerevan, 375036, Armenia, CIS}
\bigskip\bigskip

Abstract: A modified rigid string theory with infrared behaviour
governed by a nontrivial fixed point is presented.\bigskip

     Some years ago the authors of  the  works  [1-4]  considered  a
string theory with the action containing, besides  the  usual  Nambu
term, also a term equal to the square  of  the  extrinsic  curvature
tensor of the world surface:
$$
2\alpha_0S_0=\sqrt g g^{\alpha\beta}g^{\gamma\delta}\nabla_\alpha
\partial_\gamma X^\mu\nabla _\beta\partial_\delta X_\mu +\sqrt g
\lambda^{\alpha\beta}(\partial_\alpha X^\mu\partial_\beta X_\mu-g_{\alpha
\beta}).\eqno(1) $$
Here $\alpha_0$  is a  dimensionless  coupling
constant,  and  the  Lagrange
multiplier term just keeps the metric $g_{\alpha\beta}$ identical
to  the  induced
metric  $\partial_\alpha X^\mu\partial_\beta X_\mu$.  The  role  of
$S_0$   is,  classically,  to  prevent
(energetically) the surface from bending and to lead the theory to a
``smooth" phase with the string tension  scaling  near  the  critical
point. Such a smooth strings are expected to provide  a  describtion
of the universality classes of the  3-dimensional  Ising  model  and
4-dimensional QCD [2], thus passing through the  $c = 1$   barrier  of
``conformal matter + Liouville" - type theories. They are also useful
as the models of membranes, domain walls, etc. [1,3,4].

However, as was shown in the works [1-4], the coupling $\alpha_0$
turns
out to be asymptotically free, which means that in the infrared  the
first term in $S_0$  is irrelevant, while in the second one the Lagrange
multiplier field $\lambda^{\alpha\beta}$
developes an expectation value of the order  of
the cutoff, thus generating a nonzero effective string  tension  and
driving the theory into familiar (and, from the point of view of  $3d$
Ising model and $4d$ QCD, unacceptable)  Liouville  phase.  The  whole
picture is analogous, apart  from  a  numerical  difference  in  the
coefficient in the $\beta$-function, $d/2$ instead of $d-2$, to  what  happens
in non-linear sigma-models, where the quantum fluctuations  generate
a mass gap.

This situation can be avoided by generalising the action (1) in
such a way that the $\beta$-function $\beta(\alpha_0)$ has  a  sero  at  some
nonzero
value of $\alpha_0$. Then, at the fixed point,  the  continuum  theory  will
remain massless, with long-ranged correlations  between  the  normal
vectors of the surface. Such a generalisation will be  described  in
the present letter. The details and some  farther  results  will  be
presented elsewhere  [8].  The  action  to  be  presented  below  is
strongly suggested by the experience gained in attempts to  describe
the $3d$ Ising model in terms of the fermionic strings [5,6], see also
[7]. A basic object in our approach is a composite matrix field $\Omega\in
SO(d)$ [5,6], constructed in the following way.

Let $X^\mu (\xi)$ $(\mu = 1,\ldots,d)$ describe an embedding of some  (closed,
orientable)  surface $\Sigma$  into  $d$-dimensional  Euclidean  space.  The
tangent vectors of the surface will be denoted by $X_\alpha =
\partial_\alpha X$, $\alpha=1,2$, and $h_{\alpha\beta}=X_\alpha X_\beta$
is  the  2-dimensional  metric  tensor  induced  by  the
embedding. Introduce the zweibeins $e_\alpha^a$ $(a=1,2)$,  such  that
$e_\alpha^a e_{a\beta}=
X_\alpha X_\beta$, then $X_a  = e^\alpha_a X_\alpha$  are the  orthonormal
vectors
tangent  to  the surface; let us  introduce  also  $d-2$  orthonormal  normals
to  the
surface, $X_i$   $(i  =  1,\ldots,d-2)$,  and  denote  the  whole  set  of
$d$
orthonormal $d$-dimensional vectors by $X_m^\mu   =  \{ X^\mu_\alpha,X_i^\mu
\}$,
$ m  =  1,\ldots,d$. Finally, consider a matrix field $\Omega$ representing the
$SO(d)$ - rotation taking the basis set $X_m$  to some arbitrary constant
orthonormal
basis $X_m^c$ .

Some comments are now in  order.   First,  a  theory  with  the
action depending on $\Omega$ can not, in general, be  considered  a  string
theory, because of the ambiguity in the definition of $\Omega:$ it  depends
on the choice of the zweibeins $e^a_\alpha$ and of the normals $X_i$.  To  have
 a
string theory one has to ensure an $SO(2)\times SO(d-2)$ gauge  symmetry  of
the action, thus obtaining a theory with $\Omega$ taking values effectively
in the Grassmanian $G_{2,d}= SO(d)/SO(2)\times SO(d-2)$. This can be done in a
covariant way  which we will not describe here; the  net  result  is
that after a gauge fixing one ends up with the following constraints
to be imposed on $\Omega:$
$$\eqalign{e_a^{\alpha}\nabla_{\bar z}e_{b\alpha}&=0\cr
X_i^{\mu}\nabla_{\bar z}X_{\mu j}&=0\cr}\eqno(2)$$
where the covariant derivative and  the  complex  stucture  must  be
taken compatible to the metric $h_{\alpha\beta}$.

The second comment is that for some embeddings the field  $\Omega$  is
singular. This  happens  when  the  surface  has  an  open  line  of
self-intersection.  For  $d\geq4$  this  singularities  are  not  stable;
however, for $d=3$ they are stable and play the most essential role in
the string representation of $3d$ Ising  model  [5,6].  In  the  local
consideration presented here they can be ignored.

The third comment is that $\Omega$ can be taken in any  representation
of $SO(d)$. The one relevant for  the $3d$ Ising model is the spinor one
[5,6], but the construction works for a general case. In the present
letter we will take $\Omega$ to be a $d\times d$ matrix

$$\Omega_{mn}=X_m^cX_n.\eqno(3)$$

     Having this all  in  mind,  let  us  write  down  the  simplest
possible action, depending  on  $\Omega$,  namely  that  of  the  nonlinear
sigma-model:
$$4\alpha_0 S_0  =\sqrt{h}h^{\alpha\beta}tr\partial_\alpha\Omega
\partial_\beta\Omega^{-1}.\eqno(4)$$
Now substituting for $\Omega$ the expression (3) and using  the  conditions
(2) one shows that, when rewritten in terms  of  the  field  $X$,  the
action (4) coinsides with the action (1).
     This observation suggests generalising the action (1) by adding
a Wess-Zumino term to it. Namely, consider the action
$$2\alpha_0 S=\sqrt g g^{\alpha\beta}\nabla_\alpha
\partial^\gamma X\nabla_\beta\partial_\gamma X+\sqrt g\lambda^{\alpha\beta}
(\partial_\alpha X\partial_\beta X-g_{\alpha\beta})+
\sigma/3tr(\Omega^{-1}d\Omega)^3.\eqno(5)$$
In the last  term  of  this  expression  an  integration  over  some
3-dimensional manifold such that its  boundary  coinsides  with  the
surface $\Sigma$ is understood, and $\Omega$ denotes now an arbitrary extension
of
the field (3) on this manifold. The couplings must be  connected  by
Wess-Zumino quantisation condition
$$\sigma/\alpha_0  = in/4\pi,\quad n \in  Z.\eqno(6)$$
Let us stress again, that the constraints  (2),  which  have  to  be
added to the action, arise as a result of the gauge fixing  in  some
$SO(2)\times SO(d-2)$ - invariant theory, the explicit form of which is  not
important for us now.

In the  rest  of  this  letter  I  will  show, in the  one-loop
approximation, that at some value of the coupling  constant $a_0$   the
action (5) describes a conformally invariant theory. For  simplicity
I will restrict myself to the case $d=3$; I will  follow  closely  the
line of calculations described by Polyakov in the work [2].  Suppose
that the theory (5) is defined with the momentum space cutoff $\Omega$ and
let us  integrate  over  the  fields  with  momenta  satisfying  the
condition $\tilde\Lambda\leq |p|\leq\Lambda$.
Denoting this ``fast'' fields  by  the  superscript
``1'' on obtains for the quadratic  term  in  the  expansion  of  the
conformal gauge $(g_{\alpha\beta}=\rho\delta_{\alpha\beta})$  action  in
powers  of  the  ``fast''
fields:
$$S_2  = X_1 DX_1  - 2X_1 J - \lambda^{\alpha\alpha}_1\rho_1+(\partial^2
X)^2\rho_1^2
\eqno(7)$$
where
$$\eqalign{     D^{\mu\nu}&= \delta^{\mu\nu}(\partial^4-\partial_\alpha\lambda
^{\alpha\beta}\partial_\beta)+2\sigma\partial_\alpha
R_{\alpha\beta\gamma}^{\mu\nu}
\partial_\beta\partial_\gamma,\cr
R_{\alpha\beta\gamma}^{\mu\nu}&=[\partial_\delta X_\sigma^\mu X_\delta^\nu
\delta_{\alpha\beta}+\partial_\alpha X_\sigma^\mu
X_\beta^\nu]\varepsilon_{\sigma\gamma},\cr
J^\mu&=\partial_\alpha
X_\beta^\mu\lambda_1^{\alpha\beta}+\partial^2(\partial^2X^\mu)\rho_1+\cr
&+{\sigma\over 2}\partial_\alpha n^\mu
[3H_{\beta\alpha}\varepsilon^{\beta\gamma}+
iH_{\beta\delta}\varepsilon_{\beta\gamma}\varepsilon_{\alpha\delta}]\partial_\gamma\rho_1,\cr}$$
$H_{\alpha\beta}   = n^\mu\partial_\alpha X_{\beta\mu}$ is the extrinsic
curvature tensor, and we have
 put the
background (``slow'') value of $\rho$ equal to 1. $n^\mu$ is the (unique in
$d=3$)
unit vector normal to the surface. When deriving $S_2$  one has to  take
the variation of the matrix $\Omega$ keeping in mind the conditions (2).

One is left with a sequence of gaussian integrations which  are
performed quite straightforwardly, exept the one over $\lambda_1$  for which a
decomposition suggested by Polyakov [2] is used:
$$\lambda^{\alpha\alpha}=\zeta,$$

$$\partial_\alpha\lambda^{\alpha\beta}=\partial^2f^\beta.$$
As a result, in the leading logarithmic approximation the  effective
action turns out to be
$$\eqalign{2\alpha_0 S_{\rm
eff}&=\left[1-{\alpha_0\over2\pi}\log\left({\Lambda\over
\tilde\Lambda}\right)+{3i\sigma\alpha_0\over8\pi}\log\left({\Lambda\over\tilde\Lambda}\right)
\right]{(\partial^2X)^2\over\rho}+\cr
&+{\sigma\over3}{\rm
tr}(\Omega^{-1}d\Omega)^3+\lambda^{\alpha\beta}\partial_\alpha
X\partial_\beta
X-\left[1-{\alpha_0\over4\pi}\log\left({\Lambda\over\tilde\Lambda}
\right)\right]\lambda^{\alpha\alpha}\rho.\cr}$$
Performing an evident field  renormalisation  one  finally  obtains,
using the quantisation condition (6), the following  renormalisation
equation for the coupling $\alpha_0$:
$$\tilde\alpha_0=\alpha_0-{3\alpha_0^2\over4\pi}\left(1+{n\alpha_0\over8\pi}\right)\log
\left({\Lambda\over\tilde\Lambda}\right).\eqno(8) $$
{}From (8) we read off a $\beta$-function:
$$\beta(\alpha_0)={-3\alpha_0^2\over4\pi}\left(1+{n\alpha_0\over8\pi}\right).\eqno(9)$$
For $n=0$ this expression coinsides with the one found  in  the  works
[1-4] and  describes  an  asymptotically  free  coupling.  For  $n\not=0$,
however, the $\beta$-function (9) has a sero at
$\alpha_0 =-8\pi/n$. Since $\alpha$  must be
positive, one has to take $n<0$ and we obtain an infrared stable fixed
point.

     The formula (9) is the main result of the present work. Let  me
add some comments  to  it.  Though  (9)  was  derived  for  $d=3$,  an
analogous result holds also for arbitrary $d$, with the coefficient in
the $\beta$-function being $d/2$ instead of  the  $d-2$  for  the  usual $
 G_{2,d}$
Wess-Zumino-Novikov-Witten model. This  difference  arises,  as  was
stressed by Polyakov [2], as a result of restricting the field $\Omega$  by
an integrability condition stating that $\Omega$ must be obtainable from  a
surface. The formula (9) shows that imposing such a restriction does
not spoil the conformal properties of the Wess-Zumino-Novikov-Witten
model, so that the model (5,2) can be regarded as some reduction  of
it.

\bigskip
\noindent{\bf Acknowledgements}
\medskip\noindent
This work was reported on the Kiev conference on theoretical physics
(June, 1992). The author would like to  thank  the  organisers,  and
especially V. Shadura and
 S.  Pakuliak,  for  support  and  for  creation  of  the
stimulating scientific athmosphere. The author  is  grateful  to  A.
Belavin, An. Kavalov, A. Marshakov, A.  Mironov,  R.  Mkrtchyan, J. Petersen,
 J. Polchinski and A. Sedrakyan for the discussions of the  ideas  close
to those reported in this work. Stimulating  conversations  with
C. Bachas and J.L.F. Barbon are gratefully acknowledged.

\bigskip
\noindent{\bf References}
\medskip\noindent
[1]  L. Peliti, S. Leibler, Phys. Rev. Lett.  {\bf 54}  (1985)  1609.

\noindent
[2]  A.M. Polyakov, Nucl. Phys. B  {\bf 268} (1986) 406.

\noindent
[3]  H. Kleinert, Phys.  Lett.   {\bf 174B} (1986) 335.

\noindent
[4]  D. Foerster, Phys. Lett.  {\bf 114A} (1986) 115.

\noindent
[5]  Al. Kavalov, A.Sedrakyan, Nucl. Phys. B  {\bf 285} [FS19] (1987) 264.

\noindent
[6]  Al.   Kavalov, A.Sedrakyan, Phys. Lett.  {\bf 173B} (1986) 449.

\noindent
[7]  Al. Kavalov, I. Kostov, A. Sedrakyan, Phys. Lett.   {\bf 175B}  (1986)
     331.

\noindent
[8] Al.R. Kavalov, submitted to Nucl. Phys. B.

\end